\documentclass[a4paper,11pt]{article}
\pdfoutput=1 % if your are submitting a pdflatex (i.e. if you have
\usepackage{jheppub} % for details on the use of the package, please
\usepackage{amsmath,amssymb}
\usepackage{graphicx}
\usepackage{epsfig}
\usepackage{subfigure}

\def\gsim{\ \rlap{\raise 3pt \hbox{$>$}}{\lower 3pt \hbox{$\sim$}}\ }
\def\lsim{\ \rlap{\raise 3pt \hbox{$<$}}{\lower 3pt \hbox{$\sim$}}\ }

\def\tf{\tilde f}   %%%%%%%%%%%%%%%%%%%%%%%%
\def\th{\tilde h}   %%%%%%%%%%%%%%%%%%%%%%%%
\def\hf{\hat f}   %%%%%%%%%%%%%%%%%%%%%%%%
\def\hh{\hat h}   %%%%%%%%%%%%%%%%%%%%%%%%

\newcommand{\be}{\begin{equation}}
\newcommand{\ee}{\end{equation}}
\newcommand{\bea}{\begin{eqnarray}}
\newcommand{\eea}{\end{eqnarray}}

\newcommand{\eqn}[1]{eq.~(\ref{#1})}
\newcommand{\eqns}[2]{eqs.~(\ref{#1})-(\ref{#2})}

\renewcommand{\thefigure}{\Roman{figure}}
\newcommand{\ten}{ \mathbf{10}}

\newcommand{\sixtn}{ \mathbf{16}}

\newcommand{\hts}{ \mathbf{126}}
\newcommand{\bhts}{ \mathbf{\overline{126}}}
\newcommand{\fivesu}{\mathbf{5}}
\newcommand{\bfivesu}{ \mathbf{\overline{5}}}
\newcommand{\tensu}{ \mathbf{{10}}}
\newcommand{\btensu}{ \mathbf{\overline{10}}}
\newcommand{\fortyfsu}{ \mathbf{45}}
\newcommand{\bfortyfsu}{ \mathbf{\overline{45}}}
\newcommand{\fiftnsu}{ \mathbf{15}}

\title{Leptogenesis in  SO(10)}

\author{Chee Sheng Fong$^{a}$, }

\author{Davide Meloni$^{b}$, }

\author{Aurora Meroni$^{c}$, }

\author{Enrico Nardi$^{d}$}


\affiliation[a]{Instituto de F\'{\i}sica, 
Universidade de S\~ao Paulo,\\ C.\ P.\ 66.318, 05315-970 S\~ao Paulo, Brazil.}
\affiliation[b]{Dipartimento di Matematica e Fisica,\\
Via della Vasca Navale 84, 00146 Roma.}
\affiliation[c]{
$CP^3$-Origins \& the Danish Institute for Advanced Study Danish IAS,\\
Univ. of Southern Denmark, Campusvej 55, DK-5230 Odense.}
\affiliation[d]{
INFN, Laboratori Nazionali di Frascati,\\
  C.P. 13, 100044 Frascati, Italy.}

\emailAdd{fong@if.usp.br}
\emailAdd{meloni@fis.uniroma3.it}
\emailAdd{meroni@cp3.dias.sdu.dk}
\emailAdd{enrico.nardi@lnf.infn.it}

\abstract{We consider $SO(10)$ Grand Unified Theories (GUTs) with
  vacuum expectation values (vevs) for fermion masses in the
  $\mathbf{10} + \mathbf{\overline{126}}$ representation.  We show
  that the baryon asymmetry generated via leptogenesis is completely
  determined in terms of measured low energy observables and of one
  single high energy parameter related to the ratio of the
  $\mathbf{10}$ and $\mathbf{\overline{126}}$ $SU(2)$ doublet vevs.
  We identify new decay channels for the heavy Majorana neutrinos into
  $SU(2)$ singlet leptons $e^c$ which can sizeably affect the size of
  the resulting baryon asymmetry. We describe how to equip $SO(10)$
  fits to low energy data with the additional constraint of successful
  leptogenesis, and we apply this procedure to the fits carried out in
  ref.~\cite{Dueck:2013gca}. We show that a baryon asymmetry in
  perfect agreement with observations is obtained.  }

 \preprint{
CP3-Origins-2014-044 DNRF90 \newline
\phantom{a} \hfill DIAS-2014-44 \newline
\phantom{a} \hfill RM3-TH/14-18  
}


\begin{document}

\maketitle

\section{Introduction} 
Dark matter, neutrino masses, and the observed Baryon Asymmetry of the
Universe (BAU) are all evidences that a theory more fundamental than the
Standard Model (SM) must exist. However, so far no other observation
has been found to disagree with the SM predictions.  In particular,
precision electroweak tests, searches for rare flavour changing
processes and electric dipole moments, and the results of the first
LHC run, have all failed to find new physics, and have not provided
hints on how the present theory could be extended. Rather, Nature
seems to suggest us that the naturalness paradigm, which has guided
the construction of quite appealing extensions of the SM, might need
revision: maybe the realm of new physics is way above the electroweak
scale, and not around the corner.  If this is the case, any
information we are able to infer about the structure of the theory at
very large energy scales acquires pivotal importance. In this respect,
it is unequivocal that the three SM gauge couplings tend to converge
while flowing towards high energies, and this strongly suggests that
grand unification might be a fundamental feature of the underlying
theory, an idea which is also corroborated by the possibility of
explaining, within GUT frameworks, electric charge quantization, the
absence of SM gauge anomalies, etc.  The fact that in the SM gauge
coupling unification at a single point does not occur, should not be
considered as a problem, but rather as an expected feature, since the
three evidences for new physics mentioned above in general require new
matter fields below the GUT scale which can affect the running of the
couplings and can give rise, for the low energy observer, to a certain
amount of aberration rather than to a sharp focus point.

Among the possible GUT groups, $SO(10)$\cite{early} is particularly
interesting for several reasons: it is the smallest group for which
the fifteen fermions of one SM generation can fit within a single
irreducible representation (the spinorial $\mathbf{16}$), it predicts
the existence of one right-handed (RH) neutrino per family which in
turn, via the seesaw mechanism, can account for massive light
neutrinos, it can allow for non-supersymmetric gauge coupling
unification and for a sufficiently long nucleon lifetime~\cite{del
  Aguila:1980at,Babu:1992ia,Deshpande:1992au,Lavoura:1993vz,%
  bertolini3,Altarelli:2013aqa} and, being the group rank five, it can
allow for an intermediate scale a few order of magnitude below the GUT
scale where the gauge group reduces to rank 4.  Neutrino masses, the
mechanism generating the BAU, and possibly also dark matter, might all
be related with this scale.

Several connections between $SO(10)$ high energy parameters and
observables can be pinned down by studying the Yukawa sector.  Vacuum
expectation values (vevs) giving rise to fermion masses must belong to
conjugate representations of $\sixtn \otimes \sixtn = \ten \oplus
\mathbf{120} \oplus \hts $.  In a renormalizable model, the $\bhts$ is
in any case unavoidable since it is the only representation containing
a SM singlet, which is needed to implement the seesaw mechanism
(otherwise neutrino would have Dirac masses of the order of the
charged fermion masses). However, the minimal choice of just one
Yukawa coupling is not viable, because it is always possible to rotate
the fermionic $\sixtn$ to a basis in which the Yukawa matrix is
diagonal, with the result that the up- and down-quark masses would be
diagonal in the same basis and all the Cabibbo-–Kobayashi-–Maskawa
(CKM) mixings would thus vanish.  The possibility of $\mathbf{120}
\oplus \bhts $ was suggested in~\cite{Bajc:2005zf} but later found, by
dedicated numerical analyzes, to be not
viable~\cite{Joshipura:2011nn,Dueck:2013gca}.  The option $\mathbf{10}
\oplus \bhts $ has been instead found to allow fitting consistently
all the low energy
data~\cite{Dueck:2013gca,Joshipura:2011nn,Altarelli:2013aqa}. This can
be achieved under the assumption that the neutrino masses are
dominated by type I seesaw contributions, and after promoting the
fields in the $\mathbf{10}$ to complex
fields~\cite{Babu:1992ia}.\footnote{We refer to
  \cite{Babu:1992ia,Bajc:2005zf} for details and implications of
  complexifying the $\mathbf{10}$ while forbidding an additional
  $\sixtn\, \sixtn\, \mathbf{10}^*$ Yukawa coupling.}  Moreover, as it
was recently found in~\cite{Bertolini:2012im}, in this model both the
requirements of gauge coupling unification and of a proton lifetime
above the experimental limits can be satisfied.

% $\mathbf{10} \oplus \bhts  + \mathbf{45}$ model  (the $\mathbf{45}$ 
% is needed for the correct  breaking of  the GUT gauge symmetry)

In conclusion, the results of various studies agree on the fact that
the $SO(10)$ model with scalars in the $\mathbf{10} \oplus \bhts$ (a
$\mathbf{45}$ is also needed for the correct breaking of the GUT gauge
symmetry, however it does not contribute to the SM fermion masses) has
so far succeeded in passing a large set of phenomenological tests.  In
this work we will confront the model with one more test, namely we
will study if the $\mathbf{10} \oplus \bhts$ $SO(10)$ model is able to
account for the observed amount of the BAU via the standard mechanism
of CP violating decays of the heavy Majorana neutrinos $N$ and
leptogenesis~\cite{fu86,leptoreview}.  Our main results are that the
model is indeed compatible with BAU observations. As a byproduct, we
perform a complete disentanglement between the Yukawa coupling
matrices and the values of the vevs related to fermion masses
(something that cannot be achieved with low energy fits alone) and we
also obtain some information about the structure of the intermediate
scale particle spectrum.
Various studies related to leptogenesis in  $SO(10)$  that  rely on  different sets 
of assumptions and/or on  variations   of the minimal model  have appeared in the
 literature~\cite{Altarelli:2013aqa,Buccella:2012kc,SO10leptogenesis}.
Here we stick to the  minimal $SO(10)$  model  constrained only by  the condition 
that  the numerical values of the model parameters are such that all the low energy
observables are fitted correctly.  

We start in section 2 by deriving from the $SO(10)$ Yukawa Lagrangian
the couplings of the Majorana neutrinos $N$ to the SM fields, as well
as to other $SO(10)$ fields with intermediate scale masses, which could
provide new decay channels for the $N$'s.  In section 3 we compute the
leptogenesis CP asymmetries and we write down the relevant Boltzmann
equations. In section 4 we derive the connections between the relevant
leptogenesis parameters and quantities that can be fitted from low
energy data.  In section 5 we apply our results to sets of data points
resulting from the fits carried out in ref.~\cite{Dueck:2013gca}, and
we show that the correct amount of BAU is indeed produced.  Finally in
section 6 we recap and draw our conclusions.

\section{Yukawa couplings}

Fermions are assigned to three $\mathbf{16}_a$ spinorial
representations where $a=1,2,3$ is a generation index.  Scalars
are assigned to the (fundamental) vector representation $\ten_\mu$,
where $\mu,\nu,\dots$ are $SO(10)$ indices, and to the fifth rank
antisymmetric tensor $\hts_{\mu\nu\lambda\rho\sigma}$ (satisfying the
constraint $\hts_{\mu\nu\lambda\rho\sigma}
=i\epsilon_{\mu\nu\lambda\rho\sigma\alpha\beta\gamma\delta\epsilon}
\hts_{\alpha\beta\gamma\delta\epsilon}$).
Following refs.~\cite{Mohapatra:1979nn,Nath:2001uw} we write 
the $SO(10)$ Yukawa Lagrangian as 

\begin{eqnarray}
  \label{eq:6a}
{\cal L}&=& {\cal L}_{\ten}+{\cal L}_{\hts} \,, \\
  \label{eq:6b}
-{\cal L}_{\ten}&=&
\tilde h_{ab}\left(\sixtn_a^T BC^{-1}\Gamma_\mu\sixtn_b\right)\; \ten_\mu  \,, \\
  \label{eq:6c}
-{\cal L}_{\hts}&=&
\frac{1}{5!} \tilde f_{ab}\left(\sixtn_a^T BC^{-1}
\Gamma_\mu\Gamma_\nu\Gamma_\lambda\Gamma_\rho\Gamma_\sigma
\sixtn_b\right)\;  \bhts_{\mu\nu\lambda\rho\sigma} \,,
\end{eqnarray}
where $\Gamma_\mu$ are the matrices of the ten-dimensional Clifford
algebra, $B$ is the charge conjugation matrix for the $SO(10)$ spinor
representation, and $C$ is the charge conjugation matrix for
space-time spinors. The Yukawa matrices $\tilde h_{ab}$ and $\tilde
f_{ab}$ are $3\times 3$ complex symmetric, however by a unitary
rotation of the fermion multiplets it is always possible to
define a basis in which one of the two matrices is diagonal, with real and
positive eigenvalues (and we will later assume the basis in which
$\tilde h$ is diagonal). Note that such a transformations exhausts the
freedom for field redefinition, so that the $3 (\tilde h) +
(6+6)(\tilde f) = 15$ remaining Yukawas correspond to physical
parameters. 

Our goal is to project \eqn{eq:6a} onto multiplets of the SM gauge
group ${\cal G}_{SM}= SU(2)_L\times U(1)\times SU(3)$ keeping trace of
Clebsch-Gordan coefficients between the various terms.  These
coefficients can be derived by analyzing the symmetry reduction chain
$SO(10) \supset G_I \supset {\cal G}_{SM}$ where the intermediate
$G_I$ is any (maximal) subgroup of $SO(10)$ like ${\cal
  G}_{PS}=SU(2)\times SU(2)\times SU(4)$ or ${\cal G}_{5} =
SU(5)\times U(1)$. Clearly, the result does not depend on the
particular $G_I$ chosen and therefore, even if we have in mind $ G_I=
{\cal G}_{PS}$ as the intermediate scale symmetry group, here we will
follow the chain $SO(10) \supset {\cal G}_{5} \supset {\cal G}_{SM}$
since the Majorana neutrinos, which are of utmost relevance for our
study, are immediately singled out as the $SU(5)$ singlets
$\mathbf{1}=N$.  We need the following $SO(10)\to SU(5)$ branching
rules:
\begin{equation}
 \begin{split}
  \label{eq:branchings}
 \sixtn &= \mathbf{1} \oplus \bfivesu\oplus \tensu \,, \\
\ten &= \fivesu \oplus \bfivesu \,, \\
\bhts &= \mathbf{1} \oplus \fivesu \oplus \btensu \oplus
\mathbf{15} \oplus \mathbf{\overline{45}}\oplus \mathbf{50}\,.
 \end{split}
\end{equation}
Projecting~\eqn{eq:6b} onto  the $SU(5)$ multiplets
we obtain~\cite{Nath:2001uw}: 
\begin{equation}
  \label{eq:7b}
-{\cal L}_{\ten} \to 
-{\cal L}_{h} = i 2\sqrt{2} \,  
\tilde h_{ab}\, \left[
-\mathbf{1}_a\bfivesu_{bi} \fivesu_{H_u}^i+
\tensu^{ij}_a\bfivesu_{bi} \bfivesu^{H_d}_j
+\frac{1}{8} \epsilon_{ijklm}
\tensu^{ij}_a\tensu^{kl}_b\fivesu_{H_u}^m\right]\,, 
\end{equation}
where Latin indices $i,j,\dots =1,2,\dots 5$ are $SU(5)$ indices.
%($\tensu^{ij}$ is the rank two antisymmetric $SU(5)$ representation,
%while $\ten_\mu$ is the $SO(10)$ vector)
Since the terms in square brackets are not $a\leftrightarrow b$
symmetric, it is left understood that $\th_{ab}$ (as well as
$\tf_{ab}$ in the equation below) stands for the symmetrized coupling
$\frac{1}{2} (\th_{ab}+\th_{ba})$.  In~\eqn{eq:7b} we have introduced
labels for the scalar representations $\fivesu_{H_u}$ and
$\bfivesu^{H_d}$, in order to recall that the terms in square brackets
correspond (in this order) to the neutrino Dirac coupling to the
multiplet containing the $SU(2)$ doublet $H_u$, to the usual charged
lepton and $d$-quark couplings to the scalar multiplet containing the
down type Higgs $H_d$, and to the $u$-quark Yukawa coupling.  
The projection of \eqn{eq:6c} onto $SU(5)$ representations
yields~\cite{Nath:2001uw}:
\begin{eqnarray}
\label{eq:7c}
\nonumber
-{\cal L}_{\hts} \to -{\cal L}_{f}
&=& i\sqrt{\frac{2}{15}}
% \frac{1}{2}(f_{ab}+f_{ba})
\tilde f_{ba}
\left[-\sqrt{2} \mathbf{1}_a \mathbf{1}_b \mathbf{1}^S
-\sqrt{3} \mathbf{1}_a \bfivesu_{bi}\fivesu_{\Sigma_u}^{ i} 
+\mathbf{1}_a \tensu^{ij}_b\btensu_{ij}^{\Delta}\right. \qquad \\
\nonumber
&-&\left.\frac{1}{8\sqrt{3}} \epsilon_{ijklm} \tensu^{ij}_a\tensu^{kl}_b
\fivesu_{\Sigma_u}^{ m}
+ \tensu^{ij}_a \bfivesu_{bk}\bfortyfsu^{\Sigma_d k}_{ij}\right. \\
&-&\left.
\bfivesu_{ai}\bfivesu_{bj}\fiftnsu^{ij}
\ -\frac{1}{12\sqrt{2}} 
\epsilon_{ijklm} \tensu^{lm}_a \tensu^{rs}_b \mathbf{50}^{ijk}_{rs}\right]\,.
\end{eqnarray}
The first three terms in square brackets involve the $SU(5)$ singlet
$\mathbf{1} = N$, and correspond respectively to the Majorana coupling
to the $SU(5)$ singlet scalar $\mathbf{1}^S$ that provides the $N$
masses, to a Dirac coupling to a second $u$-type scalar doublet
$\Sigma_u$, and to an interaction term (the $\btensu^\Delta$ contains only
charged scalars).  The terms in the second line couple quarks and
leptons to the additional $SU(2)$ scalar doublets $\Sigma_{u,d}$.  The two
terms in the third line do not involve $N$ and do not give corrections
to the charged fermion masses.  However, $\fiftnsu$ contains an
$SU(2)$ triplet, and a small vev for its neutral component is often
used to implement in $SO(10)$ the type II seesaw.  We have assumed
from the start that neutrino masses are dominated by the type I
seesaw, and thus these last two terms are  not relevant  for us 
and will be  omitted in the following equations. 
Then, the relevant  $SU(5)\to SU(2)_L\times SU(3)$ 
branching rules for projecting \eqns{eq:7b}{eq:7c} onto 
 ${\cal G}_{SM}$ multiplets are:
\begin{eqnarray}
  \label{eq:su5branchings}
\nonumber
\bfivesu &=& (\mathbf{2},\mathbf{1}) \oplus (\mathbf{1},\mathbf{\bar 3}) \,,\\
\tensu &=&  (\mathbf{1},\mathbf{1})  \oplus (\mathbf{1},\mathbf{\bar 3})   
\oplus (\mathbf{2},\mathbf{3})   \,,\\
\nonumber
\bfortyfsu &=&    (\mathbf{2},\mathbf{1}) \oplus \dots\,, 
%(\mathbf{1},\mathbf{\bar 3})
% \oplus (\mathbf{3},\mathbf{\bar 3}) \oplus  (\mathbf{1},\mathbf{3})
\end{eqnarray}
where the $\bfortyfsu $ contains additional % ${\cal G}_{SM}$
coloured multiplets that play no role in our analysis.  Written
explicitly, the embedding of the SM fermions into the $\bfivesu\oplus
\tensu$ of $SU(5)$ is:
\begin{equation}
  \label{eq:fiveten}
% \mathbf{1} = N\,, \qquad\qquad 
\bfivesu = 
\begin{pmatrix}
d^c_1\\[-2pt]d^c_2\\[-2pt]d^c_3\\[-2pt] e^-\\[-2pt] -\nu
\end{pmatrix}%_{\!\!\!L}
\,,
\qquad\qquad 
  \tensu = \frac{1}{\sqrt{2}} 
\begin{pmatrix} 
0   & u^c_3&-u^c_2& u_1  &d_1\\[-2pt]
-u^c_3&  0 &u^c_1 & u_2  &d_2\\[-2pt]
u^c_2 &-u^c_1& 0  & u_3  &d_3\\[-2pt] 
-u_1  & -u_2 & -u_3 & 0  &e^c\\[-2pt] 
-d_1  & -d_2 & -d_3 &-e^c&0
\end{pmatrix}% _{\!\!\!L}
\,,
\end{equation}
%
% %
% \begin{equation}
%   \label{eq:fiveten}
% \mathbf{1} = N\,, \qquad\qquad 
% \bfivesu = 
% \begin{pmatrix}
% d^c_1\\[-2pt]d^c_2\\[-2pt]d^c_3\\[-2pt] e\\[-2pt] -\nu
% \end{pmatrix}_{\!\!\!L},
% \qquad\qquad 
%   \tensu = \frac{1}{\sqrt{2}} 
% \begin{pmatrix} 
% 0   & u^c_3&-u^c_2& u_1  &d_1\\[-2pt]
% -u^c_3&  0 &u^c_1 & u_2  &d_2\\[-2pt]
% u^c_2 &-u^c_1& 0  & u_3  &d_3\\[-2pt] 
% -u_1  & -u_2 & -u_3 & 0  &e^c\\[-2pt] 
% -d_1  & -d_2 & -d_3 &-e^c&0
% \end{pmatrix}_{\!\!\!L}\,,
% \end{equation}
%
which also fixes the assignments for the scalars in the
$\fivesu,\,\bfivesu$ and $\btensu$ of $SU(5)$. As regards the
$\bfortyfsu$, it is contained in the reducible three index representation
antisymmetric in the upper indices $\mathbf{50}^{ij}_k =
\left(\fivesu^i\times \fivesu^j\right)_a\times \bfivesu_k $, and 
it can be singled out by subtracting the $\sum_i \mathbf{50}^{ij}_i$
trace part (that transforms as an irreducible $\fivesu$), that is by
imposing the five constraints $\sum_i \fortyfsu^i_{ij}=0$
($j=1,2\dots,5$).  The projection of \eqns{eq:7b}{eq:7c} onto ${\cal
  G}_{SM}$ multiplets is now straightforward, and yields:
\begin{eqnarray}
  \label{eq:GSM1}
 \nonumber
-{\cal L}_{Y} &=&   2i\sqrt{2}\, \tilde h_{ab} 
\left[-  N_a \ell_b H_u +  e^c_a \ell_b H_d 
+   Q_a d^c_b H_d +  Q_a u^c_b H_u 
\right] 
- i \sqrt{\frac  {2}{15}} \tilde f_{ab}   
\left[
\sqrt{2} N_a N_b S \right.
\\
% \nonumber
&&
\qquad\qquad + \left. \sqrt{3} N_a \ell_b \Sigma_u
+ \frac{1}{\sqrt{3}}Q_a u_b^c \Sigma_u -  
e^c_a \ell_b \Sigma_d  + \frac{1}{3} Q_a d^c_b \Sigma_d  
- 2 N_a e^c_b \Delta 
\right]\,,  % + \dots \qquad
\end{eqnarray}
where we have omitted several couplings to heavy (GUT scale) coloured
scalars. Let us draw the attention to the last term involving the
scalar field $\Delta$ with charge $-1$, since it is going to be
relevant in what follows.  By redefining:
\begin{equation}
  \label{eq:hf}
  h = 2i\sqrt{2}\, \tilde h\,,  \qquad  \qquad 
f = -   \frac{i}{3} \sqrt{\frac  {2}{15}}\, \tilde f\,, 
\end{equation}
we can finally rewrite \eqn{eq:GSM1}  as:
\begin{eqnarray}
    \label{eq:GSM2}
\nonumber
-{\cal L}_{Y} &=& 
3\sqrt{2}\, f_{ab}\, N_a N_b S
+ e^c_a \Big[ h_{ab} H_d - 3 f_{ab} \Sigma_d\Big] \ell_b
+ Q_a \Big[ h_{ab} H_d + f_{ab}  \Sigma_d \Big] d^c_b \qquad 
\\
&+&  Q_a\left[  h_{ab} H_u +\sqrt{3} f_{ab} \Sigma_u\right] u^c_b
-N_a \left[ h_{ab} H_u - 3\sqrt{3}\, f_{ab} \Sigma_u\right]\ell_b   
- 6 f_{ab} N_a e^c_b   \Delta\,. \qquad
\end{eqnarray}

\section{Leptogenesis Lagrangian and CP asymmetries}

The relevant couplings to compute the leptogenesis CP asymmetries and
washout scatterings can be read off \eqn{eq:GSM2} after spelling out
a few points characterizing the scenario.
\begin{itemize} 
\item The heavy RH neutrinos $N$ acquire an intermediate scale mass
  via the vev of the SM singlet scalar $S$ that sits in the
  $\bhts$. We define:
\begin{eqnarray}
  \label{eq:Nmass}
\sigma &=& \sqrt{\langle S^\dagger S \rangle} \,.
\end{eqnarray}
Under ${\cal G}_{PS}$ the $ \overline{\mathbf{126}}$ branches to
$(\mathbf{1},\mathbf{1},\mathbf{\bar{6}}) \oplus
(\mathbf{3},\mathbf{1},\tensu) \oplus (\mathbf{1},\mathbf{3},\btensu)
\oplus (\mathbf{2},\mathbf{2},\mathbf{15})$.  $S$ is the neutral
component and $SU(3)_c$ singlet of $(\mathbf{1},\mathbf{3},\btensu)$
so that $\sigma$ also breaks ${\cal G}_{PS} \to {\cal G}_{SM}$.

\item The scalar field $\Delta$ is the charge $-1$ component of the
  same multiplet, and  is expected to acquire an intermediate
  scale mass as well. We thus need to allow for the possibility $M_N >
  M_\Delta$, i.e.  that the decay channel $N \to e^c \Delta$ is
  open. The possibility of $N$ decays into $SU(2)_L$ singlet leptons
  and scalars was already studied in~\cite{Fong:2013gaa} (although not
  in relation with $SO(10)$) and it was found to be potentially
  interesting for leptogenesis. One of the reasons is that the
  specific decays into $e^c_1$, defined as the $SU(2)$ singlet lepton
  most weakly coupled to the $N$'s, would generate an asymmetry that
  remains completely decoupled from the thermal bath, and in
  particular unaffected, even indirectly, by potentially large $\ell_1
  H \leftrightarrow N $ washouts, since the Yukawa coupling relating
  $e_1^c $ and $\ell_1$ remains out of equilibrium down to
  temperatures $T\ll M_N$.
\item The scalar fields $\Sigma_{u,d}$ belong to $(\mathbf{2},
  \mathbf{2}, \mathbf{15})$ of ${\cal G}_{PS}$ and have a mass
  unrelated to the vev $\sigma$.  Naturalness considerations then
  suggest that $M_{\Sigma} \gg M_N $ in which case the decays $N \to
  \ell \Sigma_u$ are kinematically forbidden.  Note however, that the
  neutral components of these bi-doublets will acquire induced vevs
  proportional to the EW vevs residing in the $\ten$, via the coupling
  $(\bhts\,\hts)\,(\hts\,\ten)$~\cite{Babu:1992ia}. These induced vevs
  are of fundamental importance to achieve the correct fermion mass
  relations.
\item In general, $SO(10)$ fits to SM observables produce spectra with
  heavy Majorana neutrino masses in the range $10^{9}$--$10^{12}\,$GeV
  \cite{Dueck:2013gca,Joshipura:2011nn,Altarelli:2013aqa,Buccella:2012kc}.  
  The appropriate regime to study leptogenesis is then the
  three flavour regime~\cite{flavour0,flavour1,flavour2}, which thus
  requires considering the flavoured CP asymmetries and
  flavour-dependent washouts.
\end{itemize}

As we have already mentioned, at the unbroken $SO(10)$ level one can always
choose a basis in which one of the two matrices of Yukawa coupling is
diagonal with real non-negative entries, and we choose $h_{ab}= \hat h_a
\delta _{ab}$, while $f$ remains a generic complex symmetric matrix.
After intermediate $SO(10)\to {\cal G}_{PS}$ breaking, the appropriate
basis for leptogenesis is the basis of the $N$'s mass eigenstates,
defined via a rotation of the Majorana fields with a unitary matrix $W$
that brings the matrix $f$ in the first term in \eqn{eq:GSM2} into
diagonal form: $\hat f = W f W^T$. In this basis, the Lagrangian 
terms relevant for leptogenesis can be written as: 
\begin{equation}
  \label{eq:leptoL}
  -{\cal L}_{LG} =
\frac{1}{2} M_{N_j} N_j N_j 
-N_j\, (W \hat h)_{j\alpha}\, \ell_\alpha \, H_u 
- 6\, N_j\, (\hat f W^*)_{j\alpha}\, e^c_\alpha\,   \Delta\,, 
\end{equation}
where  
\begin{equation}
\label{eq:MN} 
M_N= 6\sqrt{2}\, \hat f\, \sigma\,.   
\end{equation}
Since at the leptogenesis temperatures $T\sim M_N \ll
M_\Sigma$ scatterings between leptons and $\Sigma_d$ do not occur, the
flavour basis $\{\ell_\alpha,e^c_\alpha\}$ remains fixed by the
leptons Yukawa interactions with the light $d$-type Higgs: $ \hat
h_{\alpha}\, e^c_\alpha\ell_\alpha \, H_d $\,.  Here and below we 
denote with Latin subscripts $j,k,\dots$ the heavy neutrinos mass
eigenstates ordered from light $(j=1)$ to heavy $(j=3)$, and with Greek
subscripts $\alpha,\beta,\dots$ the lepton flavours, ordered according
to the strength of their Yukawa couplings $\hat h_1 < \hat h_2 < \hat
h_3$. It is worth  noticing at this point, that due to the
contribution of the $f$ couplings to the lepton masses after EW
symmetry breaking, the SM mass eigenstates $e\,,\mu\,, \tau$ will not
in general coincide with the leptogenesis flavour eigenstates
$\ell_1,e^c_1;\;\ell_2,e^c_2;\; \ell_3,e^c_3$.

\subsection{CP violating asymmetries}
\label{sec:leptoCP}

The CP violating asymmetries for $N_j$ decays into leptons 
of flavour $\alpha$ are defined in terms of decay widths in the usual way:
\begin{eqnarray}
  \label{eq:CPH}
  \epsilon_{j\alpha}^H &=&  \frac{1}{\Gamma^{N_j}_{\rm tot} }
\left(\Gamma^{N_j}_{\ell_\alpha H_u} - \Gamma^{N_j}_{\bar\ell_\alpha H^*_u} 
\right) \,,\\
  \label{eq:CPD}
  \epsilon_{j\alpha}^\Delta &=&  \frac{-1}{\Gamma^{N_j}_{\rm tot}} 
\left(\Gamma^{N_j}_{e^c_\alpha \Delta} - \Gamma^{N_j}_{\bar e^c_\alpha \Delta^*} 
\right) \,,
\end{eqnarray}
where superscripts (subscripts) denote initial (final) decay states,
and
\begin{eqnarray}
  \label{eq:5}
  \Gamma^{N_j}_{\rm tot} &=& \frac{M_{N_j}}{16\pi} D_j \,, \\
D_j &\equiv& 2 \sum_\alpha\left|W_{j\alpha}\right|^2 \hat h_\alpha^2 
+  36  \hat f_j^2\,, 
\end{eqnarray}
where the factor of two takes into account gauge multiplicities of the
$SU(2)$ doublets, while $6^2=36$ originates from the prefactor in
\eqn{eq:leptoL} for the $SU(2)$ singlets term.  Note that the
asymmetry in \eqn{eq:CPD} is defined with an overall minus sign, that
compensates the fact that in the production of $e^c$ lepton number is
{\it decreased} by one unit. In defining the total width $
\Gamma^{N_j}_{\rm tot}$ \eqn{eq:5} we have neglected $W_R$ mediated
three body decays $N \to e^c \bar u^c d^c$. 
Given that all the $N$'s
are strongly coupled to at least one lepton flavour (one Yukawa
coupling is in fact related to the top Yukawa coupling) this is
justified as long as $M_{W_R}\gsim M_N$.

Computation of the CP asymmetries can be carried out in the usual
way~\cite{Covi:1996wh}.  In particular, in spite of the presence of
the new decay channel $N \to e^c \Delta$, there are no new loop
contributions. This is because $f$-related couplings appear in the
loops in the combination $\sum_\alpha (\hat f W^*)_{j\alpha}(\hat f
W^*)^\dagger_{\alpha k} \propto \delta_{jk}$ which vanishes for $k\neq
j$.  For the same reason, the total CP asymmetries in $N_j\to e^c_\alpha \Delta$
decays summed over flavours also vanish $\sum_\alpha
\epsilon^\Delta_{j\alpha}=0$. Thus, the contribution of this channel
is of the ``purely flavoured leptogenesis'' (PFL)
type~\cite{AristizabalSierra:2009bh}. For the CP asymmetries we obtain:
\begin{eqnarray}
\label{eq:epsilonH}
\epsilon_{j\alpha}^{H}\! & = &\! 
\frac{4\,\hh^2_\alpha}{16 \pi D_j} \sum_{\beta,k\neq j}
\hh^2_\beta \, {\rm Im}  \left[
W_{j\alpha}W^*_{k\alpha}
\left(W_{j\beta}W^*_{k\beta} \, g^{SV}\left(x_{kj}\right)
+W_{k\beta}W^*_{j\beta} \,g^{S'}\left(x_{kj}\right)
\right)\right]   
\,,   \\
\label{eq:epsilonD}
\epsilon_{j\alpha}^{\Delta}\! & = &\! 
\frac{72\,\hf_j}{16 \pi D_j}\sum_{\beta,k\neq j}
\hf_k \hh_\beta^2\, {\rm Im}
\left[ W^*_{j\alpha} W_{k\alpha} 
\left(W_{j\beta}W^*_{k\beta}\,g^{S}\left(x_{kj}\right) 
+W_{k\beta}W^*_{j\beta}\,g^{S'}\left(x_{kj}\right)\right) 
\right]   \,,    
\end{eqnarray}
where $x_{kj}= \frac{M_{k}^{2}}{M_{j}^{2}}= 
\frac{\hf_{k}^{2}}{\hf_{j}^{2}}$. 
The function $g^{SV}=g^S+g^V$ is the sum of the  
self energy and vertex loop functions:
\begin{eqnarray}
\label{eq:fS}
g^{S} & = &  \frac{\sqrt{x}}{1-x}\to - 
\frac{1}{\sqrt{x}}-\frac{1}{x^{3/2}}+\dots \,, \\
\label{eq:fV}
g^{V}&=& \sqrt{x}\left[1-(1+x)\ln\frac{1+x}{x}\right]\to - 
\frac{1}{2\sqrt{x}}+ \frac{1}{6\, x^{3/2}}+\dots \,,
\end{eqnarray}
where the limiting expressions hold for $x\to\infty$.  Let us note
that for the $\epsilon_{j\alpha}^{\Delta}$ asymmetries there is no
vertex contribution $g^V$.  The second term in
\eqn{eq:epsilonH}  (which is sometimes denoted as the `lepton
number conserving term') involves the self energy function
\begin{equation}
  \label{eq:fSp}
  g^{S'}(x) = \frac{1}{1-x}\to -\frac{1}{x} -\frac{1}{x^{2}}+\dots \,,
\end{equation}
and it does not contribute to the total asymmetry
$\epsilon_{j}^{H}=\sum_\alpha\epsilon_{j\alpha}^{H}$. In fact by
summing over flavour one obtains $\left|(W
  \hh^2W^\dagger)_{jk}\right|^2$ which is real. We can conclude that
the $g^{S'}$ contribution is also of the PFL
type~\cite{Covi:1996wh,flavour2}.  It is interesting to note that
while in $\epsilon_{1\alpha}^{H}$ this term is in any case subdominant, 
given that (in the hierarchical limit) $g^{S'}(x_{k1})/g^{S}
(x_{k1})\sim M_1/M_k \ll 1$, it represents instead the dominant
contribution to the $N_1$ CP asymmetry
$\epsilon_{1\alpha}^{\Delta}$. This is because at the leading order in
$M_1/M_k= \hf_1/\hf_k$, which comes from the $g^S$ expansion,
$\epsilon_{1\alpha}^{\Delta}$ vanishes, so that $g^{S}$ contributes
only at order $(\hf_1/\hf_k)^3$. Thus the first order contribution
from $g^{S'}\sim (\hf_1/\hf_k)^2$ is the dominant one.  The fact that
$\epsilon_{1\alpha}^{\Delta}$ vanishes at first order in the expansion
can be seen by extending the sum over $k$ in the first term 
of \eqn{eq:epsilonD} to include also $k=1$ (the added term is real and
does not contribute).  At leading order we obtain
\begin{equation}
  \label{eq:fk}
  \sum_k \hf_k W_{k\alpha}W^*_{k\beta} g^S(x_{k1})\quad \to \quad 
-\hf_1\, \sum_k W_{k\alpha}W^*_{k\beta} = -\hf_1 \delta_{\alpha\beta}\,. 
\end{equation}
Thus the first term in~\eqn{eq:epsilonD} is, at leading order, real
$\propto |W_{j\alpha}|^2$ and gives no contribution to
$\epsilon^\Delta_{1\alpha}$ (for $\epsilon^\Delta_{2,3}$ there is no
analogous result because the expansions \eqns{eq:fS}{eq:fSp} do not
hold).  This is yet another example of the various cancellations that
follow as a consequence of the $SO(10)$ Yukawa structure.

\subsection{Boltzmann Equations} 

We are now ready to write the network of Boltzmann Equations (BE) for
$SO(10)$ leptogenesis.  In full generality, we need three BE for the
evolution of the $N_{1,2,3}$ number densities and three for the usual
anomaly free lepton charges $\Delta_\alpha=\Delta B/3-\Delta
L_\alpha$~\cite{flavour0}.  There is, however, another quantity that
in the limit of vanishing $N$ couplings is conserved: it corresponds
to the $U(1)_{e_1}$ generator of phase transformations for the $SU(2)$
singlet lepton field $e_1$ defined as the one with the smallest
coupling to the Higgs. Since $e_1$ is an $SU(2)$ singlet it does not
participate in sphaleron processes. At the relevant temperatures  
$T \sim M_N$ (with $10^9\,{\rm GeV} \lsim M_N \ll M_\Sigma$) its
interactions with the Higgs field $H_d$ are completely out of equilibrium, 
and there are also no interactions with the massive scalars $\Sigma_d$ 
which have already disappeared from the thermal bath.
Then in the effective theory governing this regime we can set $
\hh_1^{(e_1)}, \hf_1^{(e_1)} \to 0$~\cite{Fong:2010bv}.  This results in the $U(1)_{e_1}$
invariance and in a fourth conserved charge $\Delta_4 = \Delta e_1$.
Let us thus define the following charge densities (normalized to the
entropy density) that are violated just by $N$'s interactions while
are conserved by all the reactions which are in thermal equilibrium
(including non-perturbative sphaleron processes):
\begin{eqnarray}
\label{eq:DeltaCharges}
Y_{\Delta_1} &=& \frac{1}{3}Y_{\Delta B} - 2 Y_{\Delta \ell_1}\,, \\
Y_{\Delta_2} &=& \frac{1}{3}Y_{\Delta B} - \left(2 Y_{\Delta \ell_2}+Y_{\Delta e_2}\right) \,,\\
Y_{\Delta_3} &=& \frac{1}{3}Y_{\Delta B} - \left(2 Y_{\Delta \ell_3}+Y_{\Delta e_3}\right) \,,\\
Y_{\Delta_4} &=& Y_{\Delta  e_1} \,.
\end{eqnarray}
To write down the network of flavoured BE 
% for the $(Y_\Delta)_q$ charges 
as a closed system, one needs to express the density
asymmetries of the five leptons $Y_{\Delta l} \equiv \{ Y_{\Delta
  \ell_1}, Y_{\Delta \ell_2},Y_{\Delta \ell_3}, Y_{\Delta e_2},
Y_{\Delta e_3} \}$ and of the up-type Higgs $Y_{\Delta H_u}$, which
weight the washout terms, as linear combinations of the four
$(Y_\Delta)_q$, that is:
 \begin{eqnarray}
\label{eq:AC} 
\left(Y_{\Delta l}\right)_p &=& A_{pq} \,,
\left(Y_{\Delta}\right)_q \\
Y_{\Delta H_u} &=& C_q \left(Y_{\Delta}\right)_q\,. 
\end{eqnarray}
Note in passing that while $H_u$ develops an asymmetry, there is
no asymmetry for the scalar $\Delta$.  This is because the $\Delta$
are produced in decays for which the total asymmetry vanishes, and in
addition there are no other scatterings in the plasma involving the $
\Delta$ and SM or other particles with intermediate scale masses.
Therefore the $\Delta$ are not subject to any chemical potential
equilibrium condition that could result in $Y_{\Delta_\Delta}\neq 0$.

The $5\times 4$ $A$ matrix and the $C$-vector in~\eqn{eq:AC} can be
obtained by solving the system of linear constraints corresponding to
in-equilibrium spectator reactions (Yukawa related scatterings, EW and
QCD sphalerons~\cite{Nardi:2005hs}) and exactly conserved quantities
(hypercharge).  This yields:
\begin{equation}
  \label{eq:Amatrix}
  A=\begin{pmatrix}
 -\frac{93}{220} & 
 \frac{12}{220} & 
 \frac{12}{220} & 
- \frac{2}{220} & 
\!\!\!\! \phantom{\Big|} \!\!\!   \cr
 \frac{9}{240} & 
- \frac{76}{240} & 
 \frac{4}{240} & 
 \frac{6}{240} & 
 \!\!\!\! \phantom{\Big|} \!\!\!   \cr
 \frac{9}{240} & 
 \frac{4}{240} & 
- \frac{76}{240} & 
 \frac{6}{240} & 
 \!\!\!\! \phantom{\Big|} \!\!\!   \cr
 \frac{21}{264} & 
 -\frac{68}{264} & 
\frac{20}{264} & 
 -\frac{18}{264} & 
\!\!\!\! \phantom{\Big|} \!\!\!   \cr
 \frac{21}{264} & 
 \frac{20}{264} & 
- \frac{68}{264} & 
 -\frac{18}{264} &
\!\!\!\! \phantom{\Big|} \!\!\!   \cr
\end{pmatrix}\,,
\qquad 
C = \frac{1}{880} \, \left(-37,\ -52,\ -52,\ +82  \right) \,.
\end{equation}
The final $B-L$ asymmetry resulting from leptogenesis is given by the
sum $Y_{\Delta_{B-L}} = \sum_q (Y^\infty_\Delta)_q$ where
$(Y^\infty_\Delta)_q$ are obtained by integrating the BE. Finally, with
two Higgs doublets, the  relation between the  baryon asymmetry 
and $Y_{\Delta_{B-L}}$ is:
\begin{equation}
  \label{eq:B-B-L}
  Y_{\Delta B} =\frac{8}{23} Y_{\Delta_{B-L}}\,. 
\end{equation}
The BE for the evolution of the $N_j$ densities and of the  
$(Y_\Delta)_q$ charge asymmetries, considering only decays and inverse decays read:
\begin{eqnarray}
\label{eq:BEN}
\dot Y_{N_j} &=&  -  \gamma_j \left( \frac{Y_{N_j}}{Y_{N_j}^{eq}}-1\right)\,,\\
\label{eq:BE1}
\dot Y_{\Delta_1} &=&  -\sum_j \left\{
\epsilon_{j 1}^H \,
\gamma_j \, \left(\frac{Y_{N_j}}{Y_{N_j}^{eq}}-1\right)
-\frac{1}{2}
\left(
\frac{Y_{\Delta \ell_1}}{Y^{\rm eq}_f} +
\frac{Y_{\Delta H_u}}{Y^{\rm eq}_b} \right)
\gamma^{N_j}_{\ell_1 H}  \right\}\,,
\\
\label{eq:BE2}
\dot Y_{\Delta_2} &=& \!-\sum_j \! \left\{   
\left(\epsilon_{j2}^H+\epsilon_{j2}^\Delta\right)
\gamma_j \! \left(\frac{Y_{N_j}}{Y_{N_j}^{eq}}-1\right)
-\frac{1}{2}
\left[\left(
\frac{Y_{\Delta \ell_2}}{Y^{\rm eq}_f} +
\frac{Y_{\Delta H_u}}{Y^{\rm eq}_b} \right)
\gamma^{N_j}_{\ell_2 H}  
+
\frac{Y_{\Delta e_2}}{Y^{\rm eq}_f}
\gamma^{N_j}_{e_2 \Delta}  
\right]\right\} \!, \qquad
\\
\label{eq:BE3}
\dot Y_{\Delta_3} &=& \dot Y_{\Delta_2} (2 \leftrightarrow 3)\,,  
\\
\label{eq:BE4}
\dot Y_{\Delta_4} &=&   \sum_j\left\{
\epsilon_{j 1}^\Delta
\,
\gamma_j \, \left(\frac{Y_{N_j}}{Y_{N_j}^{eq}}-1\right)
-\frac{1}{2}
\frac{Y_{\Delta e_1}}{Y^{\rm eq}_f}
\gamma^{N_j}_{e_1 \Delta}  \right\}\,, 
\end{eqnarray}
where the time derivative is defined as $\dot Y \equiv s H z
\frac{d}{d z} Y$ with $z=M_1/T$, $H\simeq 1.66 \sqrt{g_*}\, T^2/M_P$ the
Hubble parameter and $s=2\pi^2 g_* T^3/45$ the entropy density, with
$g_*=110.75 $ the number of relativistic degrees of freedom including
two Higgs doublets.  $Y_b^{\rm eq}=2 Y_f^{\rm eq} =15/(4\pi^2 g_*)$
are the boson and fermion relativistic equilibrium number densities
normalized to the entropy density, $Y^{\rm eq}_{N_j} = n^{\rm
  eq}_{N_j}/s= 45 z_j^2 {\cal K}_2(z_j)/(2\pi^4 g_*)$ is the
equilibrium density for the non-relativistic $N$'s with $z_j = \sqrt{x_{j1}}
z$, ${\cal K}_n$ is the modified Bessel function of type $n$,
$\gamma_j = n^{\rm eq}_{N_j} \Gamma^{N_j}_{\rm tot} {\cal
  K}_1(z_j)/{\cal K}_2(z_j) $ is the thermal average of the 
total decay rate $\Gamma^{N_j}_{\rm tot}$ \eqn{eq:5}, and the
$\gamma^{N_j}_{\dots}$ are similarly thermal averages of the $N_j$ partial
decay rates.

\section{Relating leptogenesis parameters to observables}

To estimate the baryon asymmetry yield of the minimal $SO(10)$ model
we need the numerical values of the partial decay widths and CP
asymmetries, and to compute these quantities we need to know the
values of the Yukawa coupling matrices and of the vev $\sigma$ that
fixes the scale of the $N$'s masses. As we will now argue, 
the  values of these parameters can be fixed almost univocally
in terms of measured low energy observables, with only one single 
high energy parameter left free.

Let us  define for the up- and down- type Higgs doublets (denoted 
with subscripts $q=u,d$) the following vev-related quantities: 
\begin{eqnarray}
  \label{eq:vevs}
v_{q} &=& \sqrt{\langle H_{q}^\dagger H_{q} \rangle 
+\langle \Sigma_{q}^\dagger \Sigma_{q}\rangle }\,, \\ 
  t_\beta &\equiv& \tan\beta = \frac{v_u}{v_d} \,, \\ 
c_q &\equiv& \cos\theta_q= 
\frac{\sqrt{\langle H_{q}^\dagger H_{q} \rangle }}{v_q}\,, \\
s_q &\equiv& \sin\theta_q= 
\frac{\sqrt{\langle \Sigma_{q}^\dagger \Sigma_{q}\rangle}}{v_q}\,,  \\
\label{eq:tq} 
t_q &\equiv& \tan \theta_q =\frac{s_q}{c_q} \,,  
\end{eqnarray}
with 
\begin{eqnarray}
  \label{eq:GF}
  c_q^2+s_q^2  &=& 1\,, \\
  v_u^2+v_d^2 &=& v^2 = (\sqrt{2} G_F)^{-1}\,,
\end{eqnarray}
and $G_F$ the Fermi constant.  After EW symmetry breaking, the fermion
mass parameters can be read off from \eqn{eq:GSM2} by projecting the
scalar bi-doublets fields $H_{u,d}$ and $\Sigma_{u,d}$ on their vevs.
This yields:
\begin{eqnarray}
   \label{eq:Mell}
\frac{1}{v}\, M_\ell &=& c_\beta 
\left[\hh c_d  - 3\, f  \,  s_d\right] \equiv \hat H-3F
\,, \\
  \label{eq:Md} 
\frac{1}{v}\, M_d &=& c_\beta \left[\hh c_d + f \,  
s_d\right] \equiv \hat H+F
\,, \\
  \label{eq:Mu}
\frac{1}{v}\,  M_u &=& s_\beta 
\left[\hh c_u + \sqrt{3}\, f \, s_u\,e^{i\delta_u}\right] 
\equiv r\left(\hat H+s F\right)
\,, \\
  \label{eq:MnuD} 
-\frac{1}{v}\, M_D &=&  s_\beta 
\left[\hh c_u -3\,\sqrt{3}\, f \, s_u\,e^{i\delta_u}\right] 
\equiv r\left(\hat H-3 sF\right) \,,
\end{eqnarray}
with $M_{\ell,d,u}$ the charged leptons, $d$- and $u$-type quarks mass
matrices, and $M_D$ the Dirac mass matrix for the neutrinos.  On the
right-hand side (RHS) of the first equalities we have written the mass matrices keeping
distinguished the vevs and the Yukawa couplings; this is needed to
write down leptogenesis related quantities. The RHS of the second
equalities is instead written following a commonly adopted
parameterization
% first introduced in~\cite{Babu:1992ia} and later
\cite{Dutta:2004zh,Dutta:2005ni,Altarelli:2010at,%
  Joshipura:2011nn,Altarelli:2013aqa,Dueck:2013gca}, which is best
suited for $SO(10)$ fits to low energy data. In these notations the
relation with the non-diagonal (${\cal M}_N$) and diagonal  ($M_N$)
Majorana neutrino mass matrix can be written as:
\begin{equation}
  \label{eq:MNnew}
{\cal M}_N= W^\dagger M_NW^* = 6\sqrt{2}\,\sigma\,f  =
r_R^{-1} \, F \,.
\end{equation}
Thus, once $F$ is given, the unitary matrix $W$ can be determined from
the requirement that it brings $F$ into its diagonal form with real
non-negative entries.  Here we follow in particular the conventions used in
ref.~\cite{Dueck:2013gca} with $\hat H$ diagonal with 3 real
non-negative entries, $F$ complex symmetric with $6+6$ (real +
imaginary) parameters, $r$ and $r_R$ real quantities and
$s\equiv|s|e^{i\delta_s}$ complex, for a total of 19 parameters (12
real and 7 imaginary).  The low energy data set consists of 9 charged
fermion masses, 2 neutrinos mass squared differences, $3+3$ quark and
lepton mixing angles and 1 CKM phase, for a total of 18 parameters (17
real and 1 imaginary). Although the number of parameters exceeds by
one the number of constraints from the data, $SO(10)$ numerical fits
are able to determine all 19 parameters (and thus to yield also
predictions for yet unmeasured quantities like 
the Pontecorvo-Maki-Nakagawa-Sakata (PMNS) Dirac phase
and the $0\nu2\beta$ effective neutrino
mass~\cite{Joshipura:2011nn,Dueck:2013gca}). This is due in part to
the nonlinearity of the relations that connect $\{\hat H, F, r, r_R,
s\}$ to the data, and also to the fact that the number of real
observables (moduli) is sizeably larger than that of real free
parameters.  All in all, fits to low energy data are able to fix the
numerical values of the whole set $\{\hat H, F, r, r_R, s\}$.  We want
to relate these numerical quantities to our Yukawa parameters
$\hh,\hf,W$ (two diagonal real matrices and one unitary, for a total of
$3+3+(3+6)=$ 9 real + 6 imaginary parameters) and to the 4+1
vevs-related quantities $\sigma,
% e^{i\delta_\sigma} 
t_\beta, s_d, s_u e^{i\delta_u}$, defined in the following way: the
vevs $\langle H_{u,d}\rangle \propto c_{u,d}$ have one common
phase~\cite{Babu:1992ia} that can be reabsorbed redefining the
remaining ones, and the phase of $s_d$ can be absorbed into $W^*$ by
redefining $\delta_u $ and $\sigma$.  As regards $\sigma$, it is in
general complex, but its overall phase does not affect low energy data
or leptogenesis, and therefore for simplicity we take $\sigma$ to be
real.  With respect to the parameterization in the RHS of
\eqns{eq:Mell}{eq:MNnew} we have one additional real parameter, that
can be identified via the relation $F \leftrightarrow f \cdot s_d$.
While low energy data are only sensitive to the product of the
coupling matrices times the vev $\langle \Sigma_d\rangle$,
leptogenesis is directly sensitive to the values of $f$ alone, and
then it can allow to disentangle the Yukawa couplings from the vevs.
%
% From \eqns{eq:Mell}{eq:MNnew} we obtain immediately:
%
% \begin{eqnarray}
% r &=& t_\beta\, \frac{c_u}{c_d} \\
% s &=& |s|e^{i\delta_s}=\sqrt{3}\, \frac{t_u}{t_d}\, e^{i\delta_u}\\
% r_R^{-1} &=& 6\sqrt{2}\, \frac{\sigma}{c_\beta s_d}\,.
% \end{eqnarray} 
%
%
More precisely, it is the value of $t_d$ (defined in \eqn{eq:tq}) that
can tell us which part of the corrections to quark/lepton universality
is due to the $f$ Yukawas (whose size is relevant for leptogenesis)
and which part is due to the vevs of the $\bhts$ bi-doublets. We will
then take $t_d$ as the new independent variable. All the other vev
related quantities can then be written as:
\begin{eqnarray}
t_u &=& \frac{1}{\sqrt{3}}\, |s|\,{t_d}\,, \qquad\qquad
\delta_u \ =\  \delta_s\,, \\
\label{eq:tbeta}
t_\beta &=& r\frac{c_d}{c_u}= r\, 
\sqrt{\frac{1+\frac{1}{3}\,|s|^2\, t^2_d}{1+t^2_d}}
% \ \approx\ r\, \sqrt{\frac{1}{1+t^2_d}} 
\,, \\
\sigma &=& 
% \frac{1}{6\sqrt{2}}\, r_R^{-1}  \,c_\beta\, s_d = 
\frac{ r_R^{-1}}{6\sqrt{2}}\, 
\sqrt{\frac{t^2_d}{(1+r^2)+(1+\frac{1}{3} r^2|s|^2)t^2_d}}\,. 
% \\  &\approx&   
% \frac{1}{6\sqrt{2}}\, \frac{r_R^{-1}}{r} \,
% \sqrt{\frac{t^2_d}{1+\frac{1}{3} |s|^2t^2_d}} \ \approx \ 
% \frac{1}{6\sqrt{2}}\, \frac{r_R^{-1}}{r} \,\> t_d\,.
\end{eqnarray} 
Finally, for the two matrices $\hh$ and $f$ we obtain:
\begin{eqnarray}
\label{eq:htd}
\hh &=& \frac{1}{c_\beta\, c_d} \hat H = \hat H\>
\sqrt{(1+r^2)+(1+\frac{1}{3} r^2|s|^2)t^2_d}  \approx r\hat H \,, \\
\label{eq:ftd}
f   &=& F  \frac{1}{c_\beta s_d}  = 
 F  \,
\sqrt{\frac{(1+r^2)+(1+\frac{1}{3} r^2|s|^2)t^2_d}{t^2_d}}
\approx \frac{r}{t_d}\, F\,, 
\end{eqnarray}
where the approximations hold for $r^2 \gg 1$, $|s|^2\ll 1$ (which
result from numerical fits) and $t_d\lsim 1$ (which is favoured
theoretically since $t_d$ is the ratio between an induced vev $\langle
\Sigma_d\rangle$ and the EW vev $\langle H_d\rangle$). From
\eqn{eq:tbeta} we can see that in this approximation $t_\beta \approx
r$, from \eqn{eq:htd} we learn that the matrix $\hat h$ is not very
sensitive with respect to changes in $t_d$ while, in contrast, from
\eqn{eq:ftd} we see that $f$ is inversely proportional to $t_d$. This
means that at fixed values of $F$, small vevs $\langle \Sigma_d
\rangle$ imply large $f$-Yukawa couplings and this would render, among
other things, more important the `exotic' decay channel $N \to e^c
\Delta$.

\section{Results}

From the equations above it should be clear that once the values of
$\{\hat H, F, r, r_R, s\}$ are determined by fits to the low energy
observables, the CP asymmetries~\eqns{eq:epsilonH}{eq:epsilonD} and
the partial decay rates entering the BE \eqns{eq:BEN}{eq:BE4} remain
determined solely in terms of $t_d$. Note, however, that the heavy
Majorana masses $M_N$ do not depend on this parameter (given that they
do not depend on the doublets vevs) and thus the RH neutrino spectrum
is univocally fixed solely by the low energy data.  As we have already
mentioned, the RH neutrino masses obtained from the numerical fits
generally fall in the range $10^9-10^{12}\,$GeV, which is a favourable
one for leptogenesis, but it should be remarked that this type of
results always implies a certain degree of tuning.  The mass of the
lightest Majorana neutrino $N_1$ would in fact more naturally lie in a
mass range well below $10^{9}\,$GeV.  To show this, let us start by
writing the seesaw formula in a generic basis:
% in which the RH neutrino mass $M_N$ is diagonal:
%
\begin{eqnarray}
m_\nu  &=& - M_D {\cal M}_N^{-1} M_D   
% \\ \nonumber M_D  &=& V  M_D^d  V^T 
% \\ \nonumber m_\nu  &=& U m^d_\nu U^T 
\,, 
\end{eqnarray}
where $m_\nu$ and $M_D$ are respectively the light neutrino and the
Dirac mass matrix (both symmetric). Let us further define the diagonal
mass matrices $m_\nu^d$ and $M_D^d$ via (unitary) Takagi
factorizations $M_D = V M_D^d V^T $ and $m_\nu = U m^d_\nu U^T$, where
$U$ is the PMNS matrix.  From the seesaw formula we have:
\begin{equation}
{\cal M}_N^{-1}  = - M_D^{-1} m_\nu   M_D^{-1}  = 
V^* \frac{1}{M_D^d}\, {\cal W}\,  m_\nu^d\,  {\cal W}^T\,    
\frac{1}{M_D^d}  V^\dagger  \,,
\end{equation}
where we have defined ${\cal W}=V^\dagger U$.  Let us now write:
\begin{eqnarray}
\label{eq:MM}
{\cal M}_N^{-1}  ({\cal M}_N^{-1} )^\dagger 
& =&  V^* \frac{1}{M_D^d}\, {\cal W}\,  
m_\nu^d\,  {\cal W}^T \,  \frac{1}{(M_D^d)^2} 
{\cal W}^*\,  m_\nu^d \, {\cal W}^\dagger   
\frac{1}{M_D^d} V^T \,.
\end{eqnarray}
Taking   the trace of this equation % \eqn{eq:MM} 
yields:
\begin{eqnarray}
\label{eq:kappa}
\frac{\kappa}{M_1^2} & =& 
% \sum_{ij}\frac{1}{(M_{Di}^d)^{2}} \left({\cal W}  
% m^d  {\cal W}^T\right)_{ij}  
% \frac{1}{ (h^d_j)^{2}} \left({\cal W}^*  
% m^d  {\cal W}^\dagger \right)_{ji} \!\!= 
\sum_{ij} \frac{1}{(M^d_{D_i} M^d_{D_j})^{2}} \;\Big| \left({\cal W}\,  
m^d\,  {\cal W}^T\right)_{ij}\! \Big|^2\,,
\end{eqnarray}
where in the left-hand side the sum of the squares of the inverse RH neutrino
masses has been expressed as $\sum_i \frac{1}{M_i^2} =
\frac{\kappa}{M_1^2}$ with $M_1$ the mass of the lightest RH neutrino 
and $\kappa\equiv 1 + \frac {M_1^2}{M_2^2}+\frac{M_1^2}{M_3^2}$ ranging 
between $1$ (strong hierarchy: $M_1\ll M_{2,3}$) and 3 (degeneracy:
$M_1\sim M_2\sim M_3$). 
% In the second equality we have used the
% fact that ${\cal W}m_\nu^d {\cal W}^T$ is a symmetric matrix.
%
The RH side of \eqn{eq:kappa} is a sum of positive definite terms, and
is thus larger than each single term.  Let us take as dominant the one
involving $(M_{D_1}^d)^4$ at the denominator. We obtain:\footnote{A
  similar result is derived in the appendix of
  ref.~\cite{Davidson:2003cq}.}
 \begin{equation}
 \label{eq:M1limit}
 M_1 <\sqrt{\kappa}   \frac{(M^d_{D_1})^2}
{\big|\sum_i {\cal W}_{1i}^2\, m^d_i\big|}\,. 
 \end{equation}
 Approximate quark-lepton Yukawa universality suggests $M_{D_1}^d \sim
 m_{up}$. On the other hand ${\cal W}$ is the product of the PMNS
 matrix $U$ and of the matrix $V$ that, in the basis in which the
 down-type Yukawa couplings are diagonal, should be approximately
 given by the CKM matrix. This implies that ${\cal W}$ should depart
 sizeably from a pure diagonal form, and thus the denominator in
 \eqn{eq:M1limit} is bounded from below. For example, for $\big|\sum_i
 {\cal W}_{1i}^2 m^d_i\big| \approx 10^{-3}\,$eV, $\kappa \sim 3$ and
 $M_{D_1}^d\sim m_{up} \sim 4\,$MeV, the upper bound $M_1 \lsim 3
 \times 10^7\, {\rm GeV}$ is obtained.  Departures from quark-lepton
 Yukawa universality can at most yield a factor of a few enhancement
 of $M^d_{D_1}$ with respect to $m_{up}$, and therefore
 \eqn{eq:M1limit} can give an estimate of the amount of tuning
 required to enforce cancellations in $\big|\sum_i {\cal W}_{1i}^2\,
 m^d_i\big|$ such that $M_1$ can be lifted into the
 $10^9$--$10^{12}\,$GeV range. If from one hand it is somewhat
 unpleasant that the numerical values that we will use result from a
 certain amount of tuning in fitting $SO(10)$ parameters, on the other
 hand we find intriguing that, without any knowledge of what is
 required for leptogenesis to be successful, low energy data alone
 force all the $M_j$'s in the correct ballpark.

 By means of numerical integration of the BE
 (\ref{eq:BEN})-(\ref{eq:BE4}) it is now possible to verify if a set
 of fitted data points $\{\hat H, F, r, r_R, s\}$ can yield an amount
 of baryon asymmetry in agreement with observations.  We use the most
 recent combined Planck and WMAP CMB
 measurements~\cite{Ade:2013zuv,Bennett:2012zja}, which yield at
 $95\%$ c.l.
%
% [What we quote in the second review~\cite{leptoreview} is: $100
% \Omega_b h^2 = 2.258^{+0.057}_{-0.056} $, 7-years
% WMAP~\cite{Komatsu:2010fb} gives $100 \Omega_b h^2 =
% 2.2249^{+0.056}_{-0.057} $, 9-years WMAP~\cite{Hinshaw:2012aka} gives
% $100 \Omega_b h^2 = 2.2264\pm 0.00050$, Planck~\cite{Ade:2013zuv}
% gives (including WMAP7) $100 \Omega_b h^2 = 2.205\pm 0.00028
% (68\%c.l.)$, what are we using in the plots ?  Recall $Y_{\Delta B} =
% 0.389\times 10^{-10} \times (100 \Omega_B h^2$ ]
%
\begin{equation}
  \label{eq:CMB}
  Y_{\Delta B}^{\rm CMB}=(8.58\pm 0.22)\times 10^{-11}\,.  
\end{equation}
We take the $\{\hat H, F, r, r_R, s\}$ data points from the fits 
of Dueck and Rodejohann (DR)~\cite{Dueck:2013gca}  
to non-supersymmetric $SO(10)$ with scalars in $\ten+\bhts$,   
that are labeled as MN (minimal non-supersymmetric).

Starting from the low energy data set (SM fermion masses and mixings)
at the scale $\mu=M_Z$, DR perform two different types of fits to the
MN model. In the first approach, which they denote as MN-RGE, the
observable are evolved from the high energy scale down to $M_Z$,
integrating out the heavy neutrinos $N_j$ one by one at the
appropriate scale. The outcome of the running is then compared with
the experimental data. This is the most sophisticated approach, and in
particular is expected to yield a more reliable fit to the heavy
neutrino masses. This, besides having sizeable effects on the neutrino
parameters~\cite{Antusch:2002rr}, is a quite crucial ingredient in
leptogenesis.  The $N$ spectrum obtained with this procedure is:
\begin{equation}
  \label{eq:MN-RGE}
\{M_{N_1},\,   M_{N_2} ,\,  M_{N_3}\} 
= \{1.2 \times 10^{11},\, 2.0 \times 10^{11},\, 3.6 \times 10^{12}\} 
\;{\rm GeV}\,.  
\end{equation}
As regards the numerical values of the set $\{\hat H, F, r, r_R, s\}$,
they can be found in appendix~A of~\cite{Dueck:2013gca} labeled as
MN-RGE and are not recopied here. The main approximation in the DR
analysis is that of neglecting effects of the intermediate scale
states on gauge coupling unification and on the running of the Yukawa
matrices, and it is quite hard to estimate the related uncertainty on
the fitted parameters.\footnote{A preliminary tentative in this
  direction, although in a slightly different setup, has been done in
  \cite{Meloni:2014rga}, where it has been shown that threshold
  effects at the intermediate scale can produce effects on the fermion
  observables at the electroweak scale as large as 30\%.}

Since, as said
above, the details of the $N$'s mass spectrum is one of the most
influential ingredient for the outcome of leptogenesis, we will
present our results allowing for a $3\%$ fluctuation around the
central values in \eqn{eq:MN-RGE}. While we make no claim that this
fluctuation is accounting for the aforementioned theoretical
uncertainty, it can still be illustrative of the sensitivity of the
results on changes in the details of the $N$'s spectrum.

The second approach followed by DR, that they denote as MN-noRGE, is
based on a direct fit to the low energy neutrino parameters, and to
the GUT scale values of the charged fermion observables, evolved to
the high scale ignoring the effects of non-degenerate RH neutrinos.
As it is clearly explained in the DR paper, this second approach
cannot be considered fully consistent, however it provides a second
reference point for our study which allows for an important comparison
for the outcome of leptogenesis. The $N$'s mass spectrum for the
MN-noRGE case is:
\begin{equation}
  \label{eq:MN-noRGE}
\{M_{N_1},\,   M_{N_2} ,\,  M_{N_3}\} 
= \{1.5 \times 10^{10},\, 7.2 \times 10^{11},\, 5.5 \times 10^{12}\} 
\;{\rm GeV}\,, 
\end{equation}
while the full set $\{\hat H, F, r, r_R, s\}$ is again given in
appendix A of~\cite{Dueck:2013gca}.
 \begin{figure}[t!!]
% \FIGURE{
  \centering
  \includegraphics[scale=0.60]{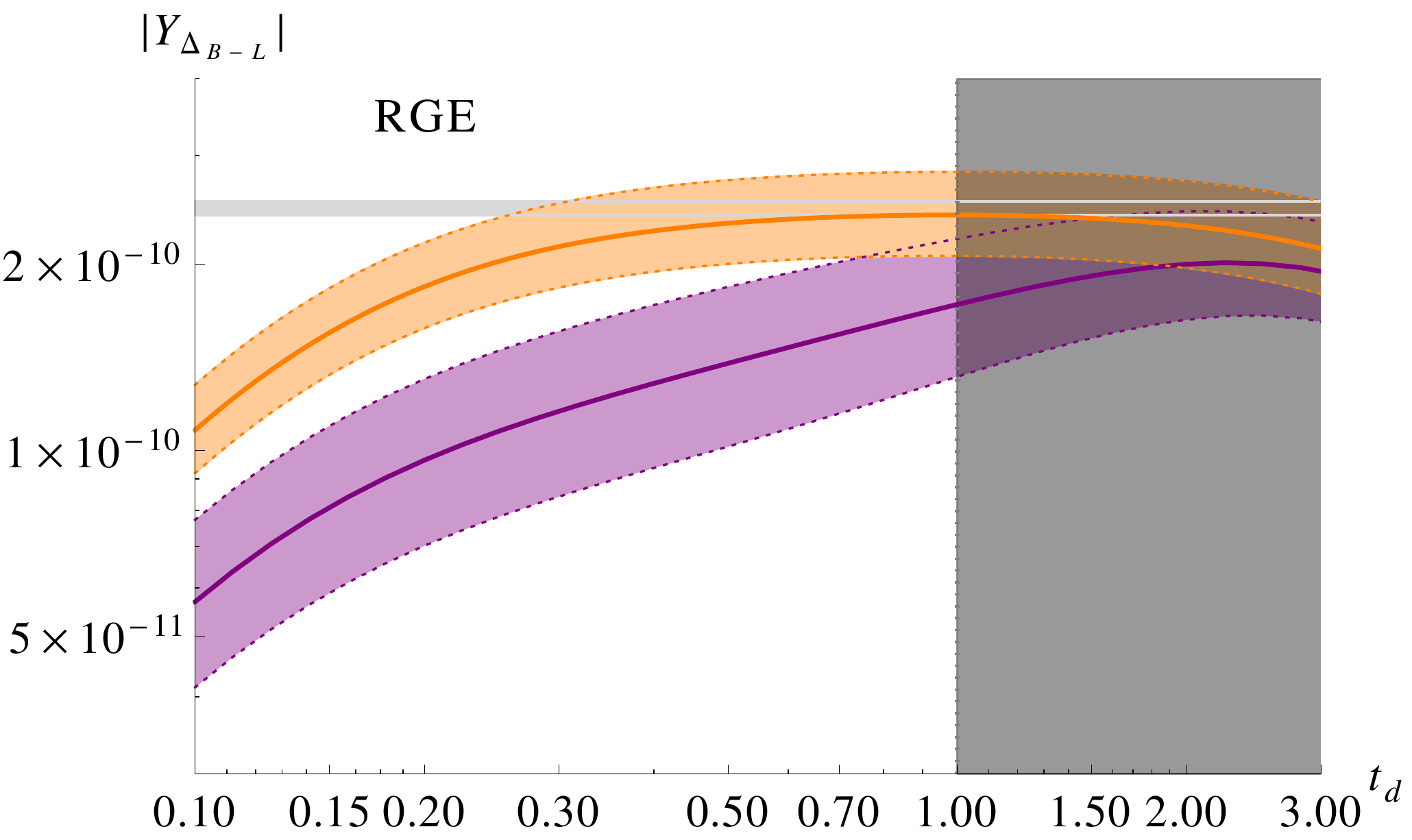}
  \vspace{0cm}
  \caption{\it The $B$-$L$ asymmetry produced with the MN-RGE data
   set of ref.~\cite{Dueck:2013gca} plotted as a function of
    $t_d$. The horizontal grey band represents the experimental limit,
    the orange curve depicts the asymmetry generated via $N_j \to
    H\ell_\alpha$ decays alone, while the purple band includes also
    the $N_j \to \Delta e^c_\alpha$ decays.  The width of the orange
    and purple bands correspond to a $3\%$  variations in the $N_j$ masses.}
  \label{fig:Delta-RGE}
% }
 \end{figure}

 Our results for the MN-RGE case are depicted in
 Fig.~\ref{fig:Delta-RGE}.  We compare the value of $Y_{\Delta_{B-L}}$ obtained
 by integrating the BE \eqns{eq:BEN}{eq:BE4}, and plotted as a
 function of the vev ratio $t_d = \Big[\frac{\langle \Sigma_d^\dagger
   \Sigma_d\rangle}{\langle H_d^\dagger H_d\rangle}\Big]^{1/2}$ with
 the experimental value which is derived from \eqn{eq:CMB} via
 \eqn{eq:B-B-L}, and that is represented by the horizontal grey band.
 The thick purple line depicts the results for the combined
 contributions of the $N_j \to \ell_\alpha H$ and $N_j \to e^c_\alpha
 \Delta$ channels and the purple band corresponds to a $3\%$ variation
 in the values of the $N_j$ masses in \eqn{eq:MN-RGE}.  We see that
 $Y_{\Delta_{B-L}}$ keeps growing with increasing values of $t_d$, that
 is with decreasing values of the $f$-couplings. The reason is that
 there is an overall contribution of the wrong sign from the $N_j \to
 e^c_\alpha \Delta$ channels which is sizeable for small $t_d$ , while
 with decreasing values of the $f$ couplings it becomes less
 important. All in all, we see that in the theoretically favoured
 region $t_d < 1$ ( i.e. $\langle \Sigma_d^\dagger \Sigma_d\rangle <
 \langle H_d^\dagger H_d\rangle$) the purple band fails to intersect
 the experimentally allowed grey region.  Only in the shaded region,
 corresponding to values $t_d > 1$, which are however theoretically
 questionable, the upper border of the purple band touches the grey
 band.  The results for the contributions of the $N_j \to \ell_\alpha
 H$ channel alone are represented by the thick orange line. This
 corresponds to the situation in which the $\Delta$ scalar is heavier
 than $N_{1,2}$ ($N_3$ contributions to leptogenesis remain quite
 marginal), that is $M_\Delta \gsim 2.0\times 10^{11}\,$GeV, so that
 $N_{1,2}$ decays into $e^c_\alpha \Delta$ are kinematically
 forbidden. In this case the predicted central value touches the
 experimental band well within the region $t_d < 1$, while the orange
 band nicely overlaps with the experimental band in the full interval
 $0.3\lsim t_d\lsim 1$.  It is also worth noticing that the maximum
 value of the $Y_{\Delta_{B-L}}$ asymmetry obtained in this model coincides
 rather precisely with the value obtained from observations.  This is
 a bit intriguing, given that a priori this value could have been
 anything.

 In summary, we find that the DR RGE fit to the minimal
 non-supersymmetric $SO(10)$ model is fully consistent with the
 requirement that the observed value of the BAU is produced via
 leptogenesis, if the two conditions (i) $M_\Delta \gsim M_{N_2}
 \simeq 2.0\times 10^{11}\,$GeV and (ii) $0.3\lsim \Big[\frac{\langle
   \Sigma_d^\dagger \Sigma_d\rangle}{\langle H_d^\dagger
   H_d\rangle}\Big]^{1/2} \lsim 1$ are satisfied.

\begin{figure}[t!]
%\FIGURE{
  \centering
  \includegraphics[scale=0.60]{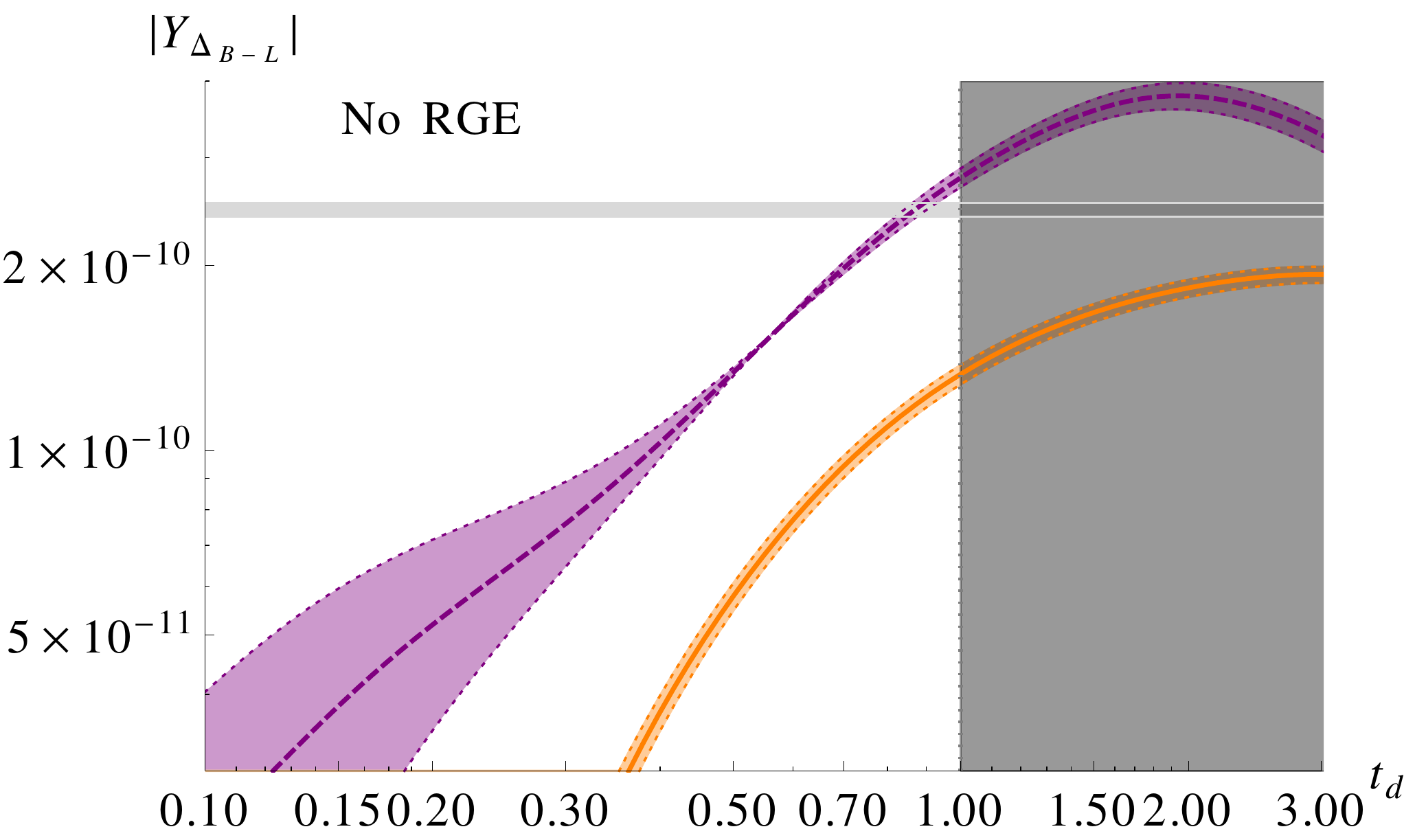}
  \vspace{0cm}
  \caption{\it Same than fig~\ref{fig:Delta-RGE} for the MN-noRGE data
    set of ref.~\cite{Dueck:2013gca}.  The thick dashed line within the
    purple band indicates a wrong sign asymmetry.}
  \label{fig:Delta-noRGE}
%} 
\end{figure}

We present for comparison in Fig.~\ref{fig:Delta-noRGE} the results
for the MN-noRGE data set.  In this case the spectrum is sufficiently
hierarchical (see~\eqn{eq:MN-noRGE}) that leptogenesis is largely
dominated by $N_1$ dynamics.  The contribution of the $N_j \to
\ell_\alpha H$ channels alone, represented by the thick orange line,
remains well below the $Y_{\Delta_{B-L}}$ experimental band even in the $t_d>1$
region. In this case the effect of the $3\%$ variations in the $N$'s
mass values is much milder than in the previous case, and this can be
traced to the larger mass hierarchy and to $N_1$ dominance.  Adding
the contributions of the $N_j \to e^c_\alpha \Delta$ channels (the
purple band and thick central line) has the striking effect of
yielding a sizeably larger asymmetry, which is however of the wrong
sign (this is depicted in the figure by means of a dashed, rather than
continuous, line).  The fact that the $N_j \to e^c_\alpha \Delta$
contributions become particularly large with increasing $t_d$ is due
to the fact that in this regime all the $N_1 \to e^c_\alpha \Delta$
decay channels enter the weak washout regime, and washout effects keep
decreasing as $t_d$ increases. $N_1 \to \ell_{1,2} H$ decays also remain in
the weak washout regime as long as $t_d \lsim 3$, while $\ell_3$ is
sufficiently strongly coupled to $N_1$ to ensure a thermal abundance
for the RH neutrino independently of initial conditions.  We also see
how the purple bands widens at small values of $t_d$ (i.e. large
values of the $f$-couplings). This is due to the vanishing of the
first order ${\cal O} (M_1/M_k)$ contribution to
$\epsilon_{1\alpha}^\Delta$ that we have discussed at the end of
section~\ref{sec:leptoCP}, which results in an enhanced sensitivity to
the $M_1/M_k$ ratio. Let us also remark at this point that the sign of
$Y_{\Delta_{B-L}}$ can be reverted (and the correct sign and size can thus be
obtained) by simply reverting the signs of all the imaginary parts of
the $F$ and $s$ parameters. While this will leave untouched the
predictions for the fermion masses and mixings, it will result in the
wrong sign for the CKM matrix $\delta_{CKM}$. Thus, it is just the
interplay between $Y_{\Delta_{B-L}}$ and $\delta_{CKM}$ which allows to rule
out the MN-noRGE data set with the contributions of the $N_j \to
e^c_\alpha \Delta$ decays included.

In summary, while the results of the DR analysis~\cite{Dueck:2013gca}
indicate that the MN-noRGE data set gives a better fit to the low
energy data than the (more reliable) MN-RGE data set, we can 
conclude that   MN-noRGE  fails to produce a sufficiently large baryon
asymmetry, and thus it does not pass the leptogenesis test.

% \noindent
% Red $= Y_{B/3-L1}$\\
% Green $= Y_{B/3-L2}$\\
% Blue $= Y_{B/3-L3}$\\
% Black $= Y_{e-e^c}$\\
% Purple $= Y_{B-L} $ \\ 

\section{Concluding remarks}

In this work we have considered leptogenesis in a non-supersymmetric
SO(10) GUT with fermion masses from the $\mathbf{10} \oplus \bhts$
Higgs representations, which can (i) fit well all the low energy data,
(ii) successfully account for unification of the gauge couplings, and
(iii) allow for a sufficiently long lifetime of the proton. We have
shown that, once the model parameters are fixed in terms of measured
low energy observables, the requirement of successful leptogenesis can
fix the only one remaining high energy parameter.  We have highlighted
that a new decay channel for the heavy Majorana neutrinos into the
$SU(2)$ singlet leptons $e^c$ is possible, and we have found that
these decays can sizeably affect the size of the resulting baryon
asymmetry. We have shown that the values of the model parameters
obtained from the fits to low energy observables given in
ref.~\cite{Dueck:2013gca} yield a baryon asymmetry in agreement with
observations.

\section*{Acknowledgments}
C. S. F.  is supported by Funda\c{c}\~ao de Amparo \`a Pesquisa do Estado de
S\~ao Paulo (FAPESP).
C. S. F. acknowledges the hospitality of Universidad de Antioquia
while this work was being completed.  A.M. acknowledges fruitful
correspondences with R. M. Syed, M. Malinsky and L. di Luzio.
D.M. acknowledges MIUR for financial support under the program Futuro
in Ricerca 2010 (RBFR10O36O).  The work of E.N. is supported in part
by the research grant "Theoretical Astroparticle Physics" number
2012CPPYP7 under the program PRIN 2012 funded by the Italian
``Ministero dell’Istruzione, Universit\'a e della Ricerca'' (MIUR) and
by the INFN ``Iniziativa Specifica'' Theoretical Astroparticle Physics
(TAsP-LNF).

\end{document}